\documentclass[aps,nofootinbib,showpacs]{revtex4}
\usepackage{graphicx}

\begin{document}

\preprint{hep-th/0311050}

\title{
Full causal dissipative cosmologies with stiff matter }

\author{M. K. Mak}
\email{mkmak@vtc.edu.hk} \affiliation{ Department of Physics, The
University of Hong Kong, Pokfulam Road, Hong Kong }

\author{T. Harko}
\email{harko@hkucc.hku.hk} \affiliation{ Department of Physics,
The University of Hong Kong, Pokfulam Road, Hong Kong }

\date{November 14, 2003}


\begin{abstract}

The general solution of the gravitational field equations for a
full causal bulk viscous stiff cosmological fluid, with bulk
viscosity  coefficient proportional to the energy density to the
power 1/4, is obtained in the flat Friedmann-Robertson-Walker
geometry. The solution describes a non-inflationary Universe,
which starts its evolution from a singular state. The time
variation of the scale factor, deceleration parameter, viscous
pressure, viscous pressure-thermodynamic pressure ratio, comoving
entropy and Ricci and Kretschmann invariants is considered in
detail.

\end{abstract}

\pacs{98.80.-k, 04.20.Jb}

\maketitle

\section{Introduction}

Dissipative bulk viscous type thermodynamical processes are
supposed to play a crucial role in the dynamics and evolution of
the early Universe. There are many processes capable of producing
bulk viscous stresses in the early cosmological fluid, like
interaction between matter and radiation, quark and gluon plasma
viscosity, different components of dark matter, particle creation,
strings and topological defects etc. Traditionally, for the
description of these phenomena the theories of Eckart \cite{Ec40}
and Landau and Lifshitz \cite{LaLi87} were used. But Hiscock and
Lindblom \cite{HiLi85} have shown that the Eckart-type theories
suffer from serious drawbacks concerning causality and stability.
Regardless of the choice of equation of state, all equilibrium
states in these theories are unstable and in addition signals may
be propagated through the fluid at velocities exceeding the speed
of light. These problems arise due to the first order nature of
the theory, i.e., it considers only first-order deviations from
the equilibrium. The neglected second-order terms are necessary to
prevent non-causal and unstable behavior.

A relativistic consistent second-order theory was found by Israel
\cite{Is76} and developed into what is called transient or
extended irreversible thermodynamics \cite{IsSt76}, \cite{HiLi89},
\cite{HiSa91}. Hiscock and Lindblom \cite{HiLi89} have shown that
these second order theories are free of the pathologies of the
Eckart-type theories. Therefore, the best currently available
theory for analyzing dissipative processes in the Universe is the
full Israel-Stewart-Hiscock causal thermodynamics.

Exact general solutions of the field equations for flat
homogeneous Universes filled with a full causal viscous fluid
source for a power-law dependence of the bulk viscosity
coefficient on the energy density have been obtained in
\cite{ChJa97}-\cite{MaHa98}. The evolution of a homogeneous and
isotropic dissipative fluid in the truncated Israel-Stewart theory
has been analyzed, by using dynamical systems methods, by Di
Prisco, Herrera and Ibanez \cite{Pr}. They have found that almost
all solutions inflate, and only a few of them can be considered
physical, since they do not satisfy the dominant energy condition.

It is the purpose of the present paper to obtain the general
solution of the gravitational field equations for the case of a
bulk viscous cosmological fluid in a flat
Friedmann-Robertson-Walker type geometry, obeying the Zeldovich
(stiff) equation of state, with the bulk viscosity coefficient
$\xi $ proportional to the energy density $\rho $ to the power
$1/4$, $\xi \propto \rho ^{1/4}$. In this case the solution of the
gravitational field equations can be represented in an exact
parametric form, in terms of Bessel functions. The behavior of the
scale factor, deceleration parameter, viscous pressure, viscous
pressure-thermodynamic pressure ratio, comoving entropy and Ricci
and Kretschmann invariants is also considered. The solution
describes a non-inflationary Universe, which starts its evolution
from a singular state.

The present paper is organized as follows. The basic equations
describing the dynamics of the viscous Friedmann-Robertson-Walker
(FRW) Universe are written down in Section II. The general
solution of the field equations is obtained in Section III. In
Section IV we discuss and conclude our results.

\section{Geometry, field equations and consequences}

The energy-momentum tensor for a bulk viscous cosmological fluid
is given by
\begin{equation}\label{1}
T_{i}^{k}=\left( \rho +p+\Pi \right) u_{i}u^{k}-\left( p+\Pi
\right) \delta _{i}^{k},
\end{equation}
where $\rho $ is the energy density, $p$ the thermodynamic
pressure, $\Pi $ the bulk viscous pressure and $u_{i}$ the four
velocity satisfying the condition $u_{i}u^{i}=1$. We use units so
that $8\pi G=c=1$.

For a flat Friedmann-Robertson-Walker Universe with a line element
given by $ds^{2}=dt^{2}-a^{2}(t)\left( dx^{2}+dy^{2}+dz^{2}\right)
$, filled with a bulk viscous cosmological fluid, the
gravitational field equations and the continuity equation
$T_{i;k}^{k}=0$ (with $;$ denoting the covariant derivative with
respect to the metric) imply
\begin{equation}\label{field}
3H^{2}=\rho , 2\dot{H}+3H^{2}=-p-\Pi , \dot{\rho}+3\left( \rho
+p\right)H=-3H\Pi ,
\end{equation}
where $H=\dot{a}/a$ is the Hubble parameter. The causal evolution
equation for the bulk viscous pressure is given by \cite{Ma95}
\begin{equation}\label{bulk}
\tau \dot{\Pi}+\Pi =-3\xi H-\frac{1}{2}\tau \Pi \left(
3H+\frac{\dot{\tau}}{\tau }-\frac{\dot{\xi}}{\xi
}-\frac{\dot{T}}{T}\right),
\end{equation}
where $T$ is the temperature, $\xi $ the bulk viscosity
coefficient and $\tau $ the relaxation time. Eq. (\ref{bulk})
arises as the simplest way (linear in $\Pi $) to satisfy the $H$
theorem ( i.e., for the entropy production to be non-negative,
$S_{;i}^{i}=\Pi ^{2}/\xi T\geq 0$, where $S^{i}=eN^{i}-\frac{\tau
\Pi ^{2}}{2\xi T}u^{i}$ is the entropy flow vector, $ e$ is the
entropy per particle and $N^{i}=nu^{i}$ is the particle flow
vector) \cite{IsSt76}, \cite{HiLi89}.

In order to close the system of equations (\ref{field}) we have to
give the equation of state for $p$ and specify $T$, $\xi $ and
$\tau $. As usual, we assume the following phenomenological laws
\cite{Ma95}:
\begin{equation}  \label{csi}
p=\left( \gamma -1\right) \rho ,\xi =\alpha \rho ^{s},T=\beta \rho
^{r},\tau =\xi \rho ^{-1}=\alpha \rho ^{s-1},
\end{equation}
where $1\leq \gamma \leq 2$, and $\alpha \geq 0$, $\beta \geq 0$,
$r\geq 0$ and $s\geq 0$ are constants. Eqs. (\ref{csi}) are
standard in cosmological models whereas the equation for $\tau $
is a simple procedure to ensure that the speed of viscous pulses
does not exceed the speed of light.

The requirement that the entropy is a state function imposes in
the present model the constraint $r=\left( \gamma -1\right)
/\gamma $ \cite{ChJa97} so that $0\leq r\leq 1/2$ for $1\leq
\gamma \leq 2$.

The growth of the total comoving entropy $\Sigma $ over a proper
time interval $\left( t_{0},t\right) $ is given by \cite{Ma95}:
\begin{equation}\label{ent}
\Sigma (t)-\Sigma \left( t_{0}\right)
=-3k_{B}^{-1}\int_{t_{0}}^{t}\Pi Ha^{3}T^{-1}dt,
\end{equation}
where $k_{B}$ is the Boltzmann constant. The
Israel-Stewart-Hiscock theory is derived under the assumption that
the thermodynamical state of the fluid is close to equilibrium,
that is the non-equilibrium bulk viscous pressure should be small
when compared to the local equilibrium pressure $\left| \Pi
\right| <<p=\left( \gamma -1\right)\rho $ \cite{Zi96}. If this
condition is violated then one is effectively assuming that the
linear theory holds also in the nonlinear regime far from
equilibrium. For a fluid description of the matter, the condition
ought to be satisfied.

To see if a cosmological model inflates or not it is convenient to
introduce the deceleration parameter $q=d\left
(H^{-1}\right)/dt-1=\left(\rho +3p+3\Pi \right)/2\rho $. The
positive sign of the deceleration parameter corresponds to
standard decelerating models, whereas the negative sign indicates
inflation.

With these assumptions the evolution equation for the flat
homogeneous causal bulk viscous cosmological models is
\begin{equation}\label{ev}
\ddot{H}+3H\dot{H}+3^{1-s}\alpha
^{-1}H^{2-2s}\dot{H}-(1+r)H^{-1}\dot{H}^{2}+\frac{9}{4}\left(
\gamma -2\right) H^{3}+\frac{3^{2-s}}{2\alpha }\gamma H^{4-2s}=0.
\end{equation}

By introducing a set of non-dimensional variables $h$ and $\theta
$ by means of the transformations $H=\alpha
_{0}h,t=\frac{2}{3\alpha _{0}}\theta $, with $\alpha _{0}=\left( \frac{3^{s}\alpha }{2}\right) ^{\frac{1}{1-2s}%
},s\neq \frac{1}{2}$, and using the expression of $r$ as a
function of $ \gamma $, Eq. (\ref{ev}) takes the form
\begin{equation}\label{evol}
\frac{d^{2}h}{d\theta ^{2}}+\left[ 2h+h^{2\left( 1-s\right) }\right] \frac{dh}{d\theta }-\left( 1+r\right) h^{-1}\left( \frac{dh}{d\theta }\right) ^{2}+%
\frac{2r-1}{1-r}h^{3}+\frac{1}{1-r}h^{2\left( 2-s\right) }=0.
\end{equation}

By denoting $n=\left( 1-2s\right) /\left( 1-r\right) $ and
changing the variables according to $h=y^{1/\left( 1-r\right)
},\eta =\int y^{1/{\left( 1-r\right) }}d\theta $, Eq. (\ref{evol})
becomes:
\begin{equation}\label{final}
\frac{d^{2}y}{d\eta ^{2}}+\left( 2+y^{n}\right) \frac{dy}{d\eta
}+\left(2r-1\right) y+y^{n+1}=0.
\end{equation}

\section{General solution for a stiff cosmological fluid with $s=1/4$}

The stiff cosmological fluid, with equilibrium pressure equal to
the energy density, is supposed to give an accurate description of
the very early phases of the evolution of the Universe. It
corresponds to the values of the parameters $\gamma =2 $ and,
consequently, $r=1/2$. We assume that the bulk viscosity
coefficient $\xi $ is of the form $\xi =\alpha \rho ^{1/4}$, with
$s=1/4$. Hence it follows that $n=1$. Therefore the nonlinear
second order ordinary differential equation (\ref{final}) becomes:
\begin{equation}\label{final1}
\frac{d^{2}y}{d\eta ^{2}}+\left( 2+y\right) \frac{dy}{d\eta
}+y^{2}=0.
\end{equation}

A particular solution of Eq. (\ref{final1}) for $n=1$, $s=1/4$ has
been obtained in \cite{MaHa00}. The general solution of Eq.
(\ref{final1}) is given by
\begin{equation}\label{finx}
y\left( \eta \right) =\frac{2C_{1}e^{-\eta }\left[ J_{1}\left(
C_{1}e^{-\eta }\right) +C_{2}Y_{1}\left( C_{1}e^{-\eta }\right)
\right] }{C_{2}Y_{0}\left( C_{1}e^{-\eta }\right) +J_{0}\left(
C_{1}e^{-\eta }\right) },
\end{equation}
where $C_{1}$ and $C_{2}$ are arbitrary integration constants,
$J_{k}\left( x\right) $ and $Y_{k}\left( x\right) $ are the Bessel
functions of the first and second kinds, respectively. They
satisfy the Bessel differential equation
\begin{equation}
x^{2}\frac{d^{2}f}{dx^{2}}+x\frac{df}{dx}+\left(x^{2}-k^{2}\right)
f=0.
\end{equation}

Hence the general solution of the gravitational field equations
for a Zeldovich causal bulk viscous fluid filled flat FRW
Universe, with bulk viscosity coefficient $\xi =\alpha \rho
^{1/4}$, can be represented in the following exact parametric
form, with $\eta $ taken as parameter:
\begin{equation}\label{sol1}
t\left( \eta \right) -t_{0}=\frac{2}{3\alpha _{0}}\int \frac{d\eta }{%
y^{2}(\eta )},H\left( \eta \right) =\alpha _{0}y^{2}\left( \eta
\right), a(\eta )=a_{0}\exp \left( \frac{2}{3}\eta \right) ,\rho
\left( \eta \right)=p\left( \eta \right) =3\alpha
_{0}^{2}y^{4}\left( \eta \right) ,
\end{equation}
\begin{equation}
q\left( \eta \right) =-\frac{3}{y\left( \eta \right)
}\frac{dy}{d\eta }-1,\Pi \left( \eta \right) =-6\alpha
_{0}^{2}y^{3}\left( \eta \right) \left[ y\left( \eta \right)
+\frac{dy}{d\eta }\right] ,\left| \frac{\Pi }{p}\right| =2\left|
1+\frac{d\ln y}{d\eta }\right| ,
\end{equation}
\begin{equation}\label{sol2}
\Sigma \left( \eta \right) =\Sigma \left( \eta _{0}\right)
+\frac{4\sqrt{3}\alpha _{0}a_{0}^{3}}{k_{B}\beta }\int y\left(
\eta \right) \left[ y\left( \eta \right) +\frac{dy}{d\eta }\right]
\exp \left( 2\eta \right) d\eta ,
\end{equation}
where $a_{0}$, $t_{0}$ and $\Sigma \left( \eta _{0}\right) $ are
constants of integration and $\alpha _{0}=2/3\alpha $.

The singular or non-singular character of the solution for all
times $t\geq 0 $ can be checked from the finite (infinite)
character of the Ricci invariant $R_{ij}R^{ij}$ and Kretschmann
scalar $R_{ijkl}R^{ijkl}$, given by
\begin{equation}
I=R_{ij}R^{ij}=12\left[ \left( \frac{\ddot{a}}{a}\right)
^{2}+\frac{\ddot{a} }{a}\left( \frac{\dot{a}}{a}\right)
^{2}+\left( \frac{\dot{a}}{a}\right) ^{4} \right] =36\alpha
_{0}^{4}y^{6}\left[ 3\left( \frac{dy}{d\eta }\right)
^{2}+3y\frac{dy}{d\eta }+y^{2}\right] ,
\end{equation}
\begin{equation}
J=R_{ijkl}R^{ijkl}=12\left[ \left( \frac{\ddot{a}}{a}\right)
^{2}+\left( \frac{\dot{a}}{a}\right) ^{4}\right] =12\alpha
_{0}^{4}y^{6}\left[ 9\left( \frac{dy}{d\eta }\right)
^{2}+6y\frac{dy}{d\eta }+2y^{2}\right] .
\end{equation}

\section{Discussions and final remarks}

The evolution of the causal bulk viscous Zeldovich fluid filled
flat Universe starts generally its evolution from a singular
state, as can be seen from the singular behavior of the invariant
$R_{ij}R^{ij}$, represented, for different values of the
integration constants $C_1$ and $C_2$, in Fig.1.

\vspace{0.2in}
\begin{figure}[h]
\includegraphics{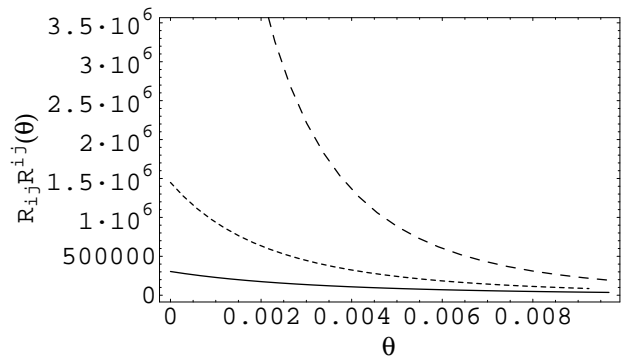}
\caption{Time evolution of the invariant $R_{ij}R^{ij}$
 for different values of the integration constants $C_1$ and $C_2$: $C_1=1$ and $C_2=-1$ (solid
curve), $C_1=1.1$ and $C_2=-0.9$ (dotted curve) and $C_1=1.1$ and
$C_2=-1$ (dashed curve).}
\label{FIG1}
\end{figure}

Generally, the dynamics of the Universe depends on the numerical
values of the constants $C_1$ and $C_2$. The behavior of the scale
factor is presented in Fig.2, for some specific values of the
integration constants. The evolution is expansionary for all
times.

\vspace{0.2in}
\begin{figure}[h]
\includegraphics{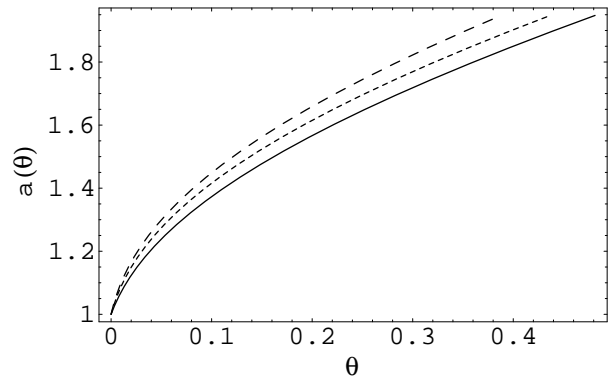}
\caption{Time variation of the scale factor $a$ for different
values of the integration constants $C_1$ and $C_2$: $C_1=1$ and
$C_2=-1$ (solid curve), $C_1=1.1$ and $C_2=-0.9$ (dotted curve)
and $C_1=1.1$ and $C_2=-1$ (dashed curve).}
\label{FIG2}
\end{figure}

The variation of the energy density of the cosmological fluid is
represented in Fig. 3.

\vspace{0.2in}
\begin{figure}[h]
\includegraphics{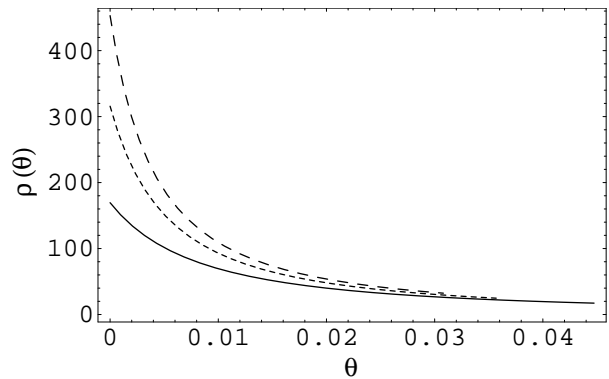}
\caption{Evolution of the energy density $\rho $ for different
values of the integration constants $C_1$ and $C_2$: $C_1=1$ and
$C_2=-1$ (solid curve), $C_1=1.2$ and $C_2=-0.8$ (dotted curve)
and $C_1=1.1$ and $C_2=-1.8$ (dashed curve). }
\label{FIG3}
\end{figure}

At the initial moment the scale factor is zero, while the energy
density tends to infinity.

The dynamics of the deceleration parameter, shown in Fig.4,
indicates, for the chosen range of the integration constant, a
non-inflationary behavior for all times, with $q>0,\forall t\geq
0$.

\vspace{0.2in}
\begin{figure}[h]
\includegraphics{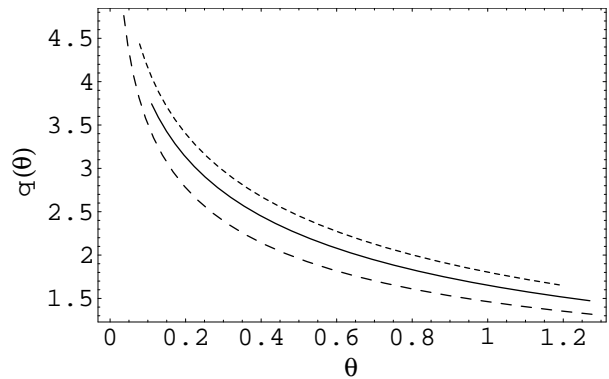}
\caption{Dynamics of the deceleration parameter $q$ for different
values of the integration constants $C_1$ and $C_2$: $C_1=1$ and
$C_2=-1$ (solid curve), $C_1=1.2$ and $C_2=-0.8$ (dotted curve)
and $C_1=1.1$ and $C_2=-1.8$ (dashed curve).}
\label{FIG4}
\end{figure}

The bulk viscous pressure $\Pi $ is negative during the
cosmological evolution, $\Pi <0,\forall t\geq 0$, as expected from
a thermodynamic point of view. In the large time limit the viscous
pressure tends to zero. In the same limit the bulk viscosity
coefficient also becomes negligible small. During the cosmological
evolution a large amount of comoving entropy is produced, with the
entropy $\Sigma $ increasing in time and tending in the large time
limit to a constant value. The time variation of the entropy is
presented in Fig. 5.

\vspace{0.2in}
\begin{figure}[h]
\includegraphics{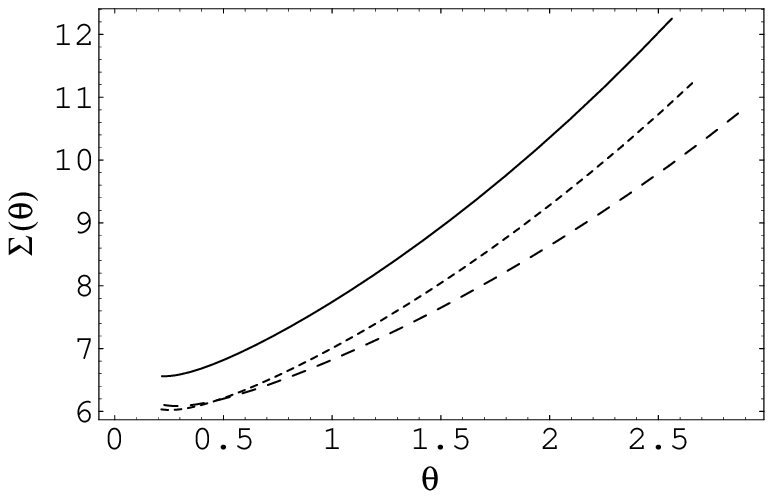}
\caption{Behavior of the comoving entropy $\Sigma $ for different
values of the integration constants $C_1$ and $C_2$: $C_1=1$ and
$C_2=-1$ (solid curve), $C_1=1.05$ and $C_2=-0.9$ (dotted curve)
and $C_1=1.08$ and $C_2=-0.8$ (dashed curve). The numerical
parameters have been normalized so that $4\sqrt{3}\alpha
_{0}a_{0}^{3}/k_{B}\beta =1$.}
\label{FIG5}
\end{figure}

The ratio of the bulk and thermodynamic pressures, $\left| \Pi
/p\right| $ is presented in Fig. 6.

\vspace{0.2in}
\begin{figure}[h]
\includegraphics{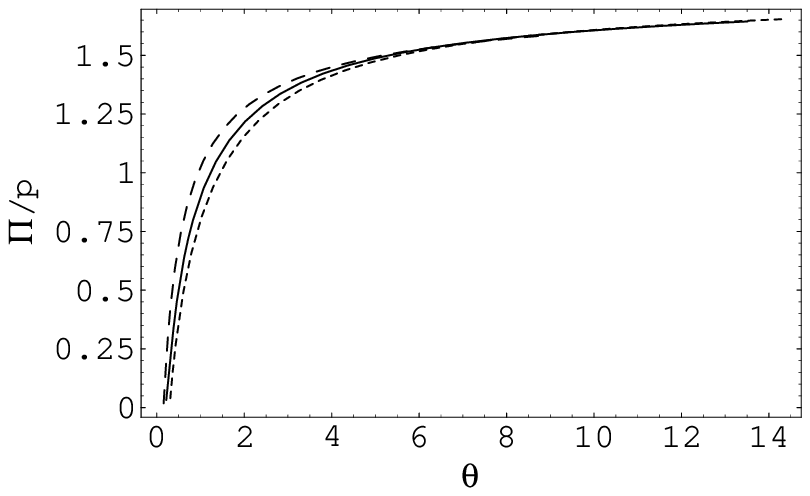}
\caption{Variation with time of the ratio $\frac{\Pi }{p}$ for
different values of the integration constants $C_1$ and $C_2$:
$C_1=1$ and $C_2=-1$ (solid curve), $C_1=1.1$ and $C_2=-0.9$
(dotted curve) and $C_1=1.2$ and $C_2=-0.9$ (dashed curve).}
\label{FIG6}
\end{figure}

For the chosen values of the integration constants the condition
$\Pi /p$  does not hold for all times, and therefore the model is
thermodynamically consistent only for a finite period of time.
Therefore the solution represented by Eqs.
(\ref{sol1})-(\ref{sol2}) can consistently describe only a
well-determined, non-inflationary period of the early dynamics of
the super-dense, matter dominated post-inflationary era, when, as
expected, the bulk viscous dissipative effects play an important
role.

\end{document}